# Inconsistent Multiple Testing Corrections: The Fallacy of Using Family-Based Error Rates to Make Inferences About Individual Hypotheses

Mark Rubin
Durham University, UK

28th March 2024

## Abstract

During multiple testing, researchers often adjust their alpha level to control the familywise error rate for a statistical inference about a joint union alternative hypothesis (e.g., "$H_{1,1}$ or $H_{1,2}$"). However, in some cases, they do not make this inference. Instead, they make separate inferences about each of the individual hypotheses that comprise the joint hypothesis (e.g., $H_{1,1}$ and $H_{1,2}$). For example, a researcher might use a Bonferroni correction to adjust their alpha level from the conventional level of 0.050 to 0.025 when testing $H_{1,1}$ and $H_{1,2}$, find a significant result for $H_{1,1}$ ($p < 0.025$) and not for $H_{1,2}$ ($p > .0.025$), and so claim support for $H_{1,1}$ and not for $H_{1,2}$. However, these separate *individual* inferences do not require an alpha adjustment. Only a statistical inference about the union alternative hypothesis "$H_{1,1}$ or $H_{1,2}$" requires an alpha adjustment because it is based on "at least one" significant result among the two tests, and so it refers to the familywise error rate. Hence, an *inconsistent correction* occurs when a researcher corrects their alpha level during multiple testing but does not make an inference about a union alternative hypothesis. In the present article, I discuss this inconsistent correction problem, including its reduction in statistical power for tests of individual hypotheses and its potential causes vis-à-vis *error rate confusions* and the *alpha adjustment ritual*. I also provide three illustrations of inconsistent corrections from recent psychology studies. I conclude that inconsistent corrections represent a symptom of *statisticism*, and I call for a more nuanced inference-based approach to multiple testing corrections.

The subject of multiple testing has received additional attention in the wake of the replication crisis. The concern is that uncorrected multiple testing is a major cause of false positive results (i.e., Type I errors) and unexpectedly low replication rates. Consequently, there is a renewed emphasis on researchers "doing the right thing" and correcting their significance thresholds (alpha levels) in order to account for inflated Type I error rates during multiple testing.

In this article, I caution that an unqualified push for multiple testing corrections may have negative consequences. In particular, I argue that it may encourage what I call *inconsistent multiple testing corrections*: adjustments to alpha levels that are inconsistent with the specific statistical inferences that are being made. To illustrate this problem, I draw attention to cases in which researchers adjust their alpha level to control family-based Type I error rates (e.g., familywise error rates) but then do not make any inferences about associated family-based hypotheses. Instead, they only make inferences about individual hypotheses, which do not require an alpha adjustment. I argue that inconsistent corrections are problematic not only logically, but also because they result in an unnecessary loss of statistical power.

To be clear, I am not opposed to an alpha adjustment for multiple testing under the appropriate circumstances. Hence, this is not an "anti-adjustment article" (Frane, 2019, p. 3). It is a pro-consistency article! My key point is that researchers should be logically consistent in their use of multiple testing corrections. If researchers use multiple testing corrections, then they should make corresponding statistical inferences about family-based *joint* hypotheses. They should not correct their alpha level and then only proceed to make statistical inferences about *individual* hypotheses because, as I explain later, such inferences do not require an alpha adjustment.

I begin by introducing the multiple testing problem and the alpha adjustment solution. I consider two common family-based error rates (the familywise error rate and the per family error rate), and I explain how associated alpha adjustments control these error rates. I then describe and illustrate inconsistent multiple testing corrections, in which a researcher adjusts their alpha level to control the error rate for a statistical inference about a family-based joint hypothesis but then only makes statistical inferences about individual hypotheses. I consider two reasons for inconsistent corrections: (a) *error rate confusions* and (b) conformity to an *alpha adjustment ritual*. I highlight recent evidence from García-Pérez (2023) showing that inconsistent corrections are likely to be common. I also explain how inconsistent corrections lead to a loss of statistical power. Finally, I illustrate my argument with three examples from recent psychology studies. I conclude that inconsistent corrections represent a symptom of the broader problem of *statisticism*, and I call for a more nuanced, inference-based approach to multiple testing corrections.

## The Multiple Testing Problem

The multiple testing problem occurs when a researcher uses more than one significance test to make a statistical inference. In this case, their Type I error rate for that inference may exceed the conventional nominal alpha level of 0.050. For example, consider a researcher who uses three significance tests to make a statistical inference about a single joint null hypothesis. Here, each of the three tests refers to a separate *constituent* null hypothesis: $H_{0,1}$, $H_{0,2}$, and $H_{0,3}$. These three constituent hypotheses comprise a *joint* hypothesis. The alpha level for determining significance with respect to each constituent hypothesis can be described as the constituent alpha level or $\alpha_{Constituent}$, and the alpha level for the final decision about rejecting or not rejecting the joint null hypothesis can be described as the joint alpha level or $\alpha_{Joint}$ (Rubin, 2021b).

If the researcher is prepared to accept a significant result on *at least one* of their three tests as sufficient grounds to reject the joint null hypothesis, then the joint null hypothesis is represented as the intersection of each of the three constituent null hypotheses: "$H_{0,1}$ and $H_{0,2}$ and $H_{0,3}$." The hypotheses are related to one another by the logical operator "and" because a significant result in relation to *any one* of them (i.e., $p < \alpha_{Constituent}$) would be sufficient to provisionally reject the entire *intersection null hypothesis* and make an inference about the corresponding *union alternative hypothesis*: "$H_{1,1}$ or $H_{1,2}$ or $H_{1,3}$." Hence, formally, this test is called a *union-intersection* test (e.g., Hochberg & Tamrane, 1987, p. 28; Kim et al., 2004; Parker & Weir, 2020, p. 563; Roy, 1953).

To provide a more concrete example, imagine that the three constituent alternative hypotheses refer to gender differences in attitudes towards biology ($H_{1,1}$), chemistry ($H_{1,2}$), and physics ($H_{1,3}$) and that the researcher is interested in making a statistical inference about a gender difference in attitudes towards these science subjects. In this case, a significant result in relation to any one of the three constituent hypotheses, in either direction (i.e., men > women or women > men), would be sufficient to reject the entire intersection null hypothesis that there is no gender difference in attitudes towards biology, chemistry, and physics and make an inference about the union alternative hypothesis that there is a gender difference in either biology, chemistry, or physics.

Note that, logically, the results of a union-intersection test only warrant a statistical inference about the associated joint hypothesis. They do not warrant statistical inferences about each of the individual constituent hypotheses (García-Pérez, 2023, p. 2; Perneger, 1998, p. 1236). For example, if the researcher obtained union-intersection test results for biology $t(326) = 2.54$, $p = 0.011$; chemistry $t(326) = 0.030$, $p = .979$; and physics $t(326) = 1.44$, $p = 0.150$, then they could only make the statistical inference that there is a significant gender difference in attitudes towards *either* biology, chemistry, *or* physics (i.e., the union alternative hypothesis). The fact that a significant gender difference is observed for biology and not for either chemistry or physics is irrelevant in the context of a union-intersection test because the test treats the three hypotheses as theoretically interchangeable constituents of the same joint hypothesis rather than as separate individual hypotheses. The principle is the same as that for a one-way ANOVA (García-Pérez, 2023): A significant result entitles us to claim that there is a significant difference between at least one pair of means, but it does not allow us to specify which pair. Of course, researchers can go on to make statistical inferences about each of the three hypotheses separately. However, these individual inferences are not based on the union-intersection test. They are based on individual tests of individual null hypotheses and, as such, they do not require an alpha adjustment (García-Pérez, 2023; Rubin, 2021b).

This last point may be a little confronting to some readers. Surely, if you conduct three individual tests, then you have a greater probability of making at least one Type I error among your set of results. Yes, you do! However, (a) this inflated familywise error rate applies to the *family* of tests, not to any *individual* test within the family; (b) you continue to have the same probability of making a Type I error in relation to *each one* of your tests; and (c) it is this latter individual error rate – $\alpha_{Individual}$ – that underwrites statistical inferences about each individual hypothesis.

To illustrate, imagine that a researcher conducts three individual tests of gender differences in attitudes towards biology, chemistry, and physics using an $\alpha_{Individual}$ of 0.050 and then concludes that there is a gender difference in relation to biology, $t(326) = 2.54$, $p = 0.011$, but not in relation to either chemistry, $t(326) = 0.030$, $p = .979$, or physics, $t(326) = 1.44$, $p = .150$. In this case, experts agree that the Type I error rate for each of these three tests is not inflated above the $\alpha_{Individual}$ of 0.050 because only one test is used to make a statistical inference (decision) about each hypothesis (Armstrong, 2014, p. 505; Cook & Farewell, 1996, pp. 96–97; Fisher, 1971, p. 206; García-Pérez, 2023, p. 15; Greenland, 2021, p. 5; Hewes, 2003, p. 450; Hitchcock & Sober, 2004, pp. 24-25; Hurlbert & Lombardi, 2012, p. 30; Matsunaga, 2007, p. 255; Molloy et al., 2022, p. 2; Parker & Weir, 2020, p. 564; Parker & Weir, 2022, p. 2; Rothman, 1990, p. 45; Rubin, 2017, pp. 271–272; Rubin, 2020a, p. 380; Rubin, 2021a, 2021b, pp. 10978-10983; Rubin, 2024; Savitz & Olshan, 1995, p. 906; Senn, 2007, pp. 150-151; Sinclair et al., 2013, p. 19; Tukey, 1953, p. 82; Turkheimer et al., 2004, p. 727; Veazie, 2006, p. 809; Wilson, 1962, p. 299). In short, if a researcher uses a single test to make a statistical inference about a single null hypothesis, then their alpha level for that inference does not become inflated and no multiple testing correction is necessary. Importantly, this principle applies even if the researcher makes millions of such individual inferences side-by-side within the same study and/or using the same dataset.

In contrast, in the case of union-intersection testing, the probability of making a Type I error about the intersection null hypothesis will always be greater than the nominal alpha level for each test ($\alpha_{Constituent}$) because the researcher has multiple opportunities to incorrectly reject the intersection null hypothesis. For example, if the intersection null hypothesis consists of three constituent null hypotheses, then the researcher will have three opportunities to make a Type I error about the intersection null hypothesis based on the three tests that they conduct using $\alpha_{Constituent}$. Hence, a multiple testing correction *is* necessary in this case in order to control the familywise error rate at the nominal level of $\alpha_{Joint}$.

In summary, multiple testing increases the probability that *at least one* of your significant results is a false positive, but it doesn't increase the probability that *each one* of your significant results is a false positive, and so if you make an inference about a joint null hypothesis that can be rejected following *at least one* significant result, then an alpha adjustment is necessary, and if you don't, then it isn't! Hence, a multiple testing correction is necessary when undertaking multiple tests of an intersection null hypothesis, but not when undertaking single tests of multiple individual null hypotheses.

## The Alpha Adjustment Solution

During union-intersection testing, the alpha adjustment solution involves lowering $\alpha_{Constituent}$ until the associated family-based error rate is less than or equal to $\alpha_{Joint}$. There are several different ways of computing the degree to which $\alpha_{Constituent}$ should be lowered, and they depend on the type of family-based error rate that is being controlled. For illustrative purposes, I consider two simple approaches that refer to the *familywise error rate* and the *per family error rate*.

### The Familywise Error Rate

The *familywise error rate* is the probability that *at least one* of the constituent test results is a Type I error (i.e., a false positive). The probability that a single constituent test yields a *true negative* (i.e., a nonsignificant result when the constituent null hypothesis is true) is $1 - \alpha_{Constituent}$. The probability that a *family* (collection) of $k$ constituent tests all yield true negatives is equal to the product of the probabilities that each yields a true negative, assuming that test results are independent of one another: $(1 - \alpha_{Constituent})^k$. Hence, the familywise error rate that *at least one* of $k$ tests yields a false positive result is $1 - (1 - \alpha_{Constituent})^k$.

Hence, if three constituent hypotheses are tested, each with an $\alpha_{Constituent}$ of 0.050, then the familywise Type I error rate will be $1 - (1 - 0.050)^3$, which equals 0.143. In this case, the familywise error rate will be greater than a nominal conventional $\alpha_{Joint}$ level of 0.050. Consequently, to control the familywise error rate at the level of $\alpha_{Joint}$, the Dunn–Šidák correction may be used to reduce $\alpha_{Constituent}$ from 0.050 to $1 - (1 - \alpha_{Constituent})^{1/3}$, which equals 0.0167. In this case, the familywise error rate will be equal to $1 - (1 - 0.0167)^3$, which equals the $\alpha_{Joint}$ level of 0.050.

**The Per Family Error Rate**

The per family error rate represents another family-based error rate. It is the number of constituent Type I errors that are expected to occur within a family of $k$ tests, and it is calculated as the sum of the $\alpha_{Constituent}$ values for each of the constituent hypotheses that are tested (Frane, 2015). Hence, if the $\alpha_{Constituent}$ values are the same for all constituent hypotheses, then the per family error rate is equal to $\alpha_{Constituent} \times k$.

For small values of $k$, the per family error rate is almost the same as the familywise error rate. However, as $k$ increases, the per family error rate becomes larger than the familywise error rate and, unlike the familywise error rate, it can become larger than 1.00. For example, if 100 constituent tests are conducted, and each has an $\alpha_{Constituent}$ of 0.050, then the familywise error rate will be 0.99 but the per family error rate will be 5.00. In other words, there will almost certainly be one or more false positive results within the family, and we should expect there to be five false positive results in total.

The Bonferroni correction may be used to control the per family error rate using the formula $\alpha_{Constituent}/k$. Hence, if $k = 3$, then the Bonferroni correction would reduce $\alpha_{Constituent}$ to 0.0169 in order to control the per family error rate at the $\alpha_{Joint}$ level of 0.050 (i.e., $0.0169 \times 3$). Note that, because the familywise error rate is the same as or smaller than the per family error rate, the Bonferroni correction may also be used to provide conservative control over the familywise error rate.

# Inconsistent Corrections

An inconsistent multiple testing correction occurs when a researcher corrects their alpha level for a union-intersection test of a joint hypothesis but then only makes statistical inferences about individual hypotheses. For example, they might correct $\alpha_{Constituent}$ in order to control a family-based error rate at the nominal conventional $\alpha_{Joint}$ of 0.050 but then only make statistical inferences about individual hypotheses, which can be made using an unadjusted conventional $\alpha_{Individual}$ of 0.050. In this case, their alpha adjustment is inconsistent with their statistical inferences about individual hypotheses, which are the only inferences that are made.

Why do researchers adjust their alpha level to control family-based error rates for family-based joint hypotheses and then fail to make statistical inferences about those hypotheses? I think there are two reasons for these inconsistent corrections: *error rate confusions* and the *alpha adjustment ritual*.

**(1) Error Rate Confusions**

Four error rate confusions may lead to inconsistent multiple testing corrections. Confusion I occurs when researchers incorrectly assume that multiple instances of individual testing somehow inflate *individual* Type I error rates for each individual inference. As previously explained, they don't! During individual testing, $\alpha_{Individual}$ refers to the probability that a single test will incorrectly reject a single hypothesis. There is no union-intersection testing in this situation, no multiple opportunities to make each Type I error, and so no error rate inflation for each statistical inference. As discussed in Confusion III below, it is true that multiple testing increases the probability of making at least one Type I error in a collection of individual tests, but it is also true that multiple testing does not increase the probability of making a Type I error with respect to each test and, during individual testing, it is only this individual Type I error rate that is relevant to researchers' statistical inferences.

Confusion II occurs when researchers incorrectly assume that multiple instances of individual testing inflate *family-based* Type I error rates for each individual inference. Again, they don't! During individual testing, $k = 1$ for each inference and so the familywise and per family error rates for each inference have the same value as the individual error rate (i.e., $\alpha_{Individual} = 1 - [1 - \alpha_{Constituent}]^1 = \alpha_{Constituent} \times 1$).

Confusion III occurs when researchers assume that multiple instances of individual testing inflate family-based error rates for *families of separate statistical inferences*. They do! However, these family-based error rates are irrelevant to *each* statistical inference! To illustrate, consider a researcher who computes the familywise error rate for 20 separate individual statistical inferences that each use an $\alpha_{Individual}$ of 0.050. In this case, the researcher assumes that $k = 20$ instead of $k = 1$ because they count the number of statistical inferences that are made (20) rather than the number of tests that are used to make each inference (1). The resulting familywise error rate (0.642) does not refer to the incorrect rejection of any specific null hypothesis (individual or joint) and so, by definition, it does not represent a Type I error rate. Nonetheless, the researcher may make the mistake of using this *hypothesis-free familywise error rate* to judge the stringency of each of their statistical inferences. This approach is flawed because the probability that at least one of 20 statistical inferences represents a Type I error (0.642) is irrelevant to the probability of incorrectly rejecting each individual null hypothesis (0.050). Indeed, the probability that at least one inference represents a Type I error can be astronomically high in large groups of inferences (e.g., in genome-wide association studies) without it affecting the probability of incorrectly rejecting *each* null hypothesis, which remains steadfast at a conventional unadjusted $\alpha_{Individual}$ of 0.050.

Finally, Confusion IV occurs when researchers assume that individual and family-based Type I error rates apply to *substantive* inferences rather than just *statistical* inferences (Meehl, 1997). They don't! In the frequentist framework, a statistical inference assumes that random sampling error is the only source of error, and a Type I error rate indicates the frequency with which this sampling error would lead to the incorrect rejection of a statistical null hypothesis during a long run of random sampling from the null population. In contrast, a substantive inference assumes that additional theoretical, methodological, and analytical errors may lead to the incorrect rejection of a substantive null hypothesis. Type I error rates do not account for these nonstatistical forms of error. Nonetheless, researchers may confuse substantive hypotheses with statistical hypotheses and erroneously apply Type I error rates and associated multiple testing corrections to their decisions about substantive hypotheses (Meehl, 1997).

These four error rate confusions may be exacerbated by the ambiguous phrasing that is sometimes used in explanations of the multiple testing problem (see also García-Pérez, 2023, pp. 2-4). For example, it is true that "multiple testing inflates the Type I error rate," but it is important to clarify what kind of "multiple testing," what kind of "Type I error rate," and what kind of hypothesis. Hence, it is more accurate to say that *union-intersection testing* inflates the *familywise error rate* for statistical inferences about *intersection null hypotheses*. Multiple individual tests do not inflate individual Type I error rates for inferences about individual null hypotheses. Nonetheless, the vague dictum that "multiple testing inflates the Type I error rate" may lead some researchers to incorrectly assume that (a) multiple testing inflates individual Type I error rates and (b) family-based error rates indicate the extent of this inflation.

Given their subtle and seductive nature, it is worth considering error rate confusions in relation to both the familywise error rate and the per family error rate. Taking the familywise error rate first, Confusion III may lead researchers to calculate a hypothesis-free familywise error rate for a collection of individual statistical inferences about individual hypotheses with a view to controlling the (uninflated) individual Type I error rate. Hence, a researcher who makes 20 statistical inferences about 20 individual hypotheses using an $α_{Individual}$ of 0.050 may erroneously conclude that their Type I error rate for *each inference* is inflated because their familywise error rate for this collection of inferences is 0.642. In fact, their Type I error rate for each inference remains at the $α_{Individual}$ level of 0.050. The researcher's erroneous conclusion is due to an inappropriate application of the familywise error rate to a collection of single tests of individual hypotheses.

Similarly, multiple testing inflates the per family error rate and not the individual Type I error rate. Again, failure to appreciate this point may lead to a misapplication of the per family error rate to statistical inferences about individual hypotheses. For example, a researcher might conduct 20 significance tests using an alpha level of 0.050 and obtain only one significant result. Given that this number of significant results matches the per family error rate, the researcher might then be tempted to assume that their significant result is more likely to be a Type I error. Again, however, this reasoning is flawed because it confuses Type I errors about individual null hypotheses with Type I errors about joint null hypotheses. The per family error rate is a family-based error rate and, as such, it is only appropriate when making inferences about family-based joint hypotheses. It is inappropriate to apply it to inferences about individual hypotheses.

In summary, family-based error rates tell us nothing about the probability of making a Type I error with respect to an individual null hypothesis. To believe that they do is to succumb to a type of ecological fallacy in which the Type I error rate for a decision about a family of hypotheses is misapplied to decisions about the individual hypotheses within that family. Family-based error rates only tell us the probability of making a Type I error with respect to family-based intersection null hypotheses.

## (2) The Alpha Adjustment Ritual

It is possible to resolve error rate confusions through logical reasoning. However, researchers do not select statistical approaches on the basis of logical reasoning per se. Sociocultural fashions and conventions are also influential, and it is here that an *alpha adjustment ritual* may come into play.

In his article *Mindless Statistics*, Gigerenzer (2004) noted that the "null ritual" of null hypothesis significance testing "has sophisticated aspects…such as alpha adjustment" (p. 588). He did not go into this issue any further. However, in my view, the alpha adjustment ritual involves the automatic adjustment of alpha levels whenever multiple testing occurs, regardless of whether statistical inferences are made about individual null hypotheses or intersection null hypotheses. This social ritual is supported by colleagues, peer reviewers, editors, journals, and so on, some of whom consider failure to conform to the ritual as one of the "seven deadly sins" of statistical practice (Kuzon et al., 1996; Millis, 2003; Popp et al., 2012).

Again, to be clear, an alpha adjustment is appropriate when making a statistical inference about an intersection null hypothesis on the basis of a union-intersection test. However, an alpha adjustment is not appropriate when making statistical inferences about multiple individual hypotheses on the basis of multiple individual tests. Hence, the problem with the alpha adjustment ritual is that it lacks nuance and sensitivity to the type of inferences that are made. In particular, it does not allow for the possibility that researchers make multiple individual statistical inferences about multiple individual hypotheses based on multiple individual tests. Researchers who follow the alpha adjustment ritual in this situation will end up making inconsistent multiple testing corrections because an alpha adjustment is in appropriate for the specific statistical inferences that they make.

In summary, statistical inferences about intersection null hypotheses require an alpha adjustment, but statistical inferences about individual null hypotheses do not, even if multiple such inferences are made within the same study and/or on the same data set. Contrary to the alpha adjustment ritual then, there are some cases of multiple testing that do not require an alpha adjustment, and unthinking adherence to the ritual may result in inconsistent multiple testing corrections.

## Inconsistent Corrections are Common

How common are inconsistent multiple testing corrections? In his recent review, García-Pérez (2023) checked 109 research articles that had used multiple testing corrections and that were published in the journals *Behavior Research Methods* and *Psychological Science* between 2021 and June 2022. He found that

> "an invariable feature of all papers was that each and all of the individual tests for which a *p* value was reported (whether with or without corrections) was interpreted individually, that is, there was an inference per test and the tests were never regarded as collectively addressing a joint intersection null hypothesis" (p. 4).

Hence, researchers used multiple testing corrections when they made statistical inferences about individual null hypotheses and not about the intersection null hypotheses to which their corrections would apply. We can conclude that, at least in García-Pérez's (2023) sample of articles, inconsistent multiple testing corrections are very common.

## Inconsistent Corrections Reduce Statistical Power

Inconsistent corrections also lead to an unjustifiable loss of statistical power. If a researcher adjusts their alpha level below its nominal level to account for multiple testing but only makes statistical inferences about individual hypotheses and not about a joint hypothesis, then they will have lowered the power of their individual tests for no good reason. Consequently, their Type I error rate will be unnecessarily low, and their Type II error rate will be unnecessarily high (García-Pérez, 2023, p. 11).

For example, imagine that a researcher wanted to make two statistical inferences about two individual hypotheses. Logically, they could use an unadjusted conventional $\alpha_{Individual}$ of 0.050 in each case. However, further imagine that the researcher followed the alpha adjustment ritual and used a Bonferroni correction to reduce their $\alpha_{Individual}$ level from 0.050 to 0.025 (i.e., $\alpha_{Individual}/k$). If they obtained *p* values of 0.010 and 0.040, then they could only reject the first null hypothesis. They would not be able to reject the second null hypothesis because their *p* value of 0.040 would be higher than their adjusted alpha level of 0.025. Of course, if they had not made this alpha adjustment, then they could have rejected their second hypothesis at the conventional alpha level of 0.050. Hence, the researcher's inconsistent correction caused a loss of statistical power and, assuming false null hypotheses, this loss of power would explain their nonsignificant result.

It is important to clarify here that researchers can set $\alpha_{Individual}$ to be lower than the conventional level of 0.050 if they wish to provide more stringent tests of their individual hypotheses (Parker & Weir, 2020, p. 564; Rubin, 2021b, p. 10984). However, this approach represents stringent alpha *specification* rather than an *adjustment* to a previously specified alpha level. Once α$_{Individual}$ has been set at a specified level (e.g., 0.050, 0.010, etc.), it should not be adjusted to account for multiple testing.

## Three Examples of Inconsistent Corrections

To better appreciate the implications of inconsistent corrections, it is helpful to consider three examples from recent research studies. To obtain these examples, I searched Google Scholar at the end of December 2023 for recent articles (2021 - 2023) in journals that had the word "psychology" in the title and that included the terms "0.025" and "0.05/2" or "0.050/2." I used the period 2021 to 2023 to demonstrate the contemporary nature of inconsistent corrections over the past three years. I used the term "psychology" in the journal title to try to restrict articles to psychology journals, although there is no reason to believe that the same issue does not occur in other disciplines. Finally, I used the terms "0.025" and "0.05(0)/2" because they are likely to be used when discussing a relatively simple Bonferroni correction to a conventional alpha level of 0.050 when $k = 2$. In this case, statistical inferences about the two individual hypotheses can be made using an α$_{Individual}$ of 0.050, and a statistical inference about the joint hypothesis can be made using an α$_{Constituent}$ of 0.025, which maintains the associated α$_{Joint}$ at 0.050. Hence, if researchers use this Bonferroni correction in a logically consistent manner, then they should make a statistical inference about a joint alternative hypothesis that encompasses the two constituent hypotheses that they test (e.g., "$H_{1,1}$ or $H_{1,2}$"). However, if they use it in a logically inconsistent manner, then they will not make a statistical inference about the joint hypothesis, and they will instead make two separate statistical inferences about two separate individual hypotheses (e.g., $H_{1,1}$ and $H_{1,2}$).

My search returned 62 results. In screening these results, I selected cases in which (a) the statistical analysis was relatively simple, (b) one of the two test results was significant at the 0.025 level (i.e., $p < 0.025$), and (c) the other test result was significant at the 0.050 level but nonsignificant at the 0.025 level (i.e., $0.025 < p < 0.050$). This third criterion allowed me to illustrate nonsignificant results that may be attributed to a loss of statistical power caused by inconsistent corrections on the assumption that the associated null hypotheses are false.

Using these criteria, I chose three examples: Prem et al. (2021, Study 1), Clemens and Grolig (2023), and Janssen et al. (2023, Experiment 2). I selected these studies because they provided relatively clear illustrations of inconsistent multiple testing corrections. Nonetheless, their selection does not imply that they are any less rigorous or credible than other studies. Indeed, given that the researchers restricted their statistical inferences to individual hypotheses, the selected studies can be viewed as providing *more* stringent tests than other studies because their alpha levels are lower than the conventional level of 0.050. My point here is only to highlight (a) the logical inconsistency in lowering the alpha level to control the familywise error rate and then only making claims about individual hypotheses and not about joint, family-based, hypotheses and (b) the potential implications arising from an associated loss in statistical power.

**Example 1: Prem et al. (2021, Study 1)**

Prem et al. (2021, Study 1) conducted a study to develop and validate a scale to measure the cognitive demands of planning, structuring, and coordinating flexible working arrangements. The researchers explained that, "when testing Hypotheses 2 through 5, the Bonferroni-corrected α was 0.05/2 = 0.025 because Hypotheses 2 through 5 each included 2 correlations" (p. 7). For example, Hypothesis 4 was that "the subscale for the planning of working places would be positively related to the availability of telework possibilities to work from home and the availability of telework possibilities to [work] from other locations outside the employer's premises" (p. 4).

> The researchers found that,
> "in line with Hypotheses 2 through 5, structuring of work tasks showed significant positive associations with decision-making autonomy and work methods autonomy; planning of working times showed significant positive associations with work scheduling autonomy and the availability of flextime; planning of working places showed significant positive associations with the availability of working from home and the availability of telework from other locations; and coordinating with others showed significant positive associations with initiated interdependence and received interdependence (compare Table 1). All of these correlations remained significant after Bonferroni correction, with the exception of the correlation between planning of working places and the availability of working from home. Thus, Hypotheses 2, 3, and 5 were fully supported, and Hypothesis 4 was partly supported" (Prem et al., 2021, p. 7).

Hence, the researchers tested four hypotheses, each referring to two correlations, and they adjusted $\alpha_{Constituent}$ to 0.025 (i.e., 0.050/2) in each case. Following this Bonferroni correction, they found support for three of the four hypotheses and partial support for Hypothesis 4, because only one of the two correlations was significant at the 0.025 level in this case.

The conclusion that Hypothesis 4 was only "partially supported" is the result of an inconsistent correction. The use of the Bonferroni correction implies that Hypothesis 4 is a union alternative hypothesis that can be fully supported following at least one significant result using an adjusted $\alpha_{Constituent}$ of 0.025. The researchers met this criterion, finding that planning of working places was significantly positively correlated with the availability of telework possibilities from other locations. Hence, logically, the researchers could have concluded that there was *full* support for Hypothesis 4. Instead, they concluded that Hypothesis 4 was only "partially supported." This conclusion suggests that they construed Hypothesis 4 as being composed of two individual hypotheses, and they would conclude that there was "full support" for Hypothesis 4 if both individual hypotheses were supported, "partial support" if only one hypothesis was supported, and "no support" if neither hypothesis was supported. However, in this case, no alpha adjustment is required because separate statistical inferences are made about each individual hypothesis, and a nonstatistical summary of these two inferences is then provided in relation to "Hypothesis 4" (i.e., "full support," "partial support," or "no support"). Hence, the researchers should have reported two significant results at the 0.050 level and then claimed full support for Hypothesis 4. Instead, they only reported one significant result at the 0.025 level and claimed partial support for Hypothesis 4. Assuming the null hypotheses were false, this substantive claim of partial support may be attributed to a lack of statistical power caused by the inconsistent correction.

I should note that correspondence with the first author of this study revealed that the decision to use a Bonferroni correction was made in response to a request from a peer reviewer (R. Prem, personal communication, January 03, 2024). Hence, at least in this case, a peer reviewer encouraged the researchers to follow the alpha adjustment ritual.

**Example 2: Clemens and Grolig (2023)**

Clemens and Grolig (2023) investigated how people would respond when they imagined that they were being interviewed by the police under either suspicion or no suspicion that they had committed an illegal act at a crime scene, but an act that was unrelated to the crime being investigated. Participants were asked to imagine that they had performed either a lawful act or an unlawful act at a bookstore in which a theft had taken place. In the lawful condition, participants looked at a book, and in an unlawful condition, they made an illegal purchase of a mobile phone. The researchers hypothesised "that unlawful act participants (vs. lawful act participants) would report…evasive strategies more frequently (hypothesis 1b)." The researchers considered two evasive strategies: (a) deception and (b) reluctant information sharing. They reported that,

> "as two evasive categories of strategies were identified, we applied a Bonferroni corrected significance level (0.05/2) of 0.025 for hypothesis 1b. The results show that unlawful (vs. lawful) act participants reported the evasive strategy to be deceptive ($\chi2(1, N = 128) = 28.038$, $p < 0.001$, $\phi = 0.47$) significantly more often, whereas no significant result was found for the evasive strategy of *reluctant information sharing* ($\chi2(1, N = 128) = 4.137$, $p = 0.042$, $\phi = 0.18$. These results are only partially in line with hypothesis 1b" (Clemens & Grolig, 2023, pp. 386-387).

Again, the researchers' conclusion that their results are "only partially in line" with their hypothesis is inconsistent with their analytical approach. The use of a Bonferroni correction implies that only one of the two tests needs to yield a significant result in order to reject the intersection null hypothesis that unlawful act participants would report neither of the evasive strategies more frequently than lawful act participants. Consistent with this criterion, the researchers found one significant result using an adjusted $\alpha_{Constituent}$ level of 0.025 ($p < 0.001$). However, instead of claiming full support for the union alternative hypothesis, they only claimed partial support. Again, this conclusion implies that the two tests were construed as single tests of two individual null hypotheses. In this case, however, both null hypotheses could be provisionally rejected using a conventional $\alpha_{Individual}$ at the unadjusted level of 0.050 ($p < 0.001$ & $p = 0.042$), and a substantive conclusion of "full support" could be reached.

**Example 3: Janssen et al. (2023, Experiment 2)**

Finally, Janssen et al. (2023, Experiment 2) investigated the effectiveness of different study strategies, focusing on the differences between *blocked* study (studying one topic at a time; e.g., AAA BBB CCC) and *interleaved* study (mixing up different topics across time; e.g., ACB BAC CBA). These researchers used a Bonferroni correction to adjust their alpha level to 0.025 during an independent samples *t*-test in which study strategy (blocked vs. interleaved) was the independent variable and (a) prospective judgments of learning and (b) actual learning outcomes were the two dependent variables. As they explained,

> "to test for significant differences, we used independent *t*-tests with a Bonferroni corrected significance level of $p < 0.025$ (i.e., 0.05/2). As expected and again consistent with Experiment 1, students who had used blocked studying made higher prospective judgments of learning ($M = 5.83$, $SD = 1.60$) than students who had used interleaved studying ($M = 5.24$, $SD = 1.82$), $t(297) = 2.95$, $p = 0.003$, Cohen's $d = 0.34$. Numerically, the actual learning outcomes were higher for the interleaved study condition ($M = 6.92$, $SD = 2.06$) than for the blocked study condition ($M = 5.83$, $SD = 1.60$). However, in contrast to our expectations, this difference was not statistically significant, $t(297) = -1.99$, $p = 0.048$, Cohen's $d = -0.23$" (Janssen et al., 2023, p. 24).

Hence, using an $\alpha_{Constituent}$ of 0.025, the researchers found a significant effect of study strategy (blocked vs. interleaved) on prospective judgments of learning ($p = 0.003$) but not on actual learning outcomes ($p = 0.048$). Following the logic of the Bonferroni correction, they could have then rejected the associated intersection null hypothesis and claimed full support for the union alternative hypothesis that study strategy affected either prospective judgments of learning or actual learning outcomes. Instead, they proceeded to make statistical and substantive inferences about each outcome variable separately. For example, they concluded that

> "both experiments replicated findings from prior research that, overall, at the group level, students reported higher effort investment and made lower judgments of learning during interleaved studying than during blocked studying (Kirk-Johnson et al., 2019; Onan et al., 2022). Yet, we only replicated the finding that students actually learned significantly more from interleaved studying than from blocked studying (as evidenced by their test performance) in Experiment 1. In Experiment 2, the difference in learning outcome, although numerically in the hypothesized direction, was not statistically significant…" (Janssen et al., 2023, p. 28).

If the authors wanted to control their Type I error rate for each decision about each individual hypothesis at 0.050, then they could have used an unadjusted $α_{Individual}$ of 0.050, rather than an adjusted $α_{Constituent}$ of 0.025. In this case, they would have decided that *both* of their test results were significant ($p$s = 0.003 & 0.048) rather than only their first result ($p$ = 0.003). An $α_{Constituent}$ of 0.025 would only be required if the authors wanted to make a decision about the intersection null hypothesis using an $α_{Joint}$ of 0.050. However, they did not consider this intersection null hypothesis. Hence, once again, this example illustrates an inconsistent multiple testing correction and a nonsignificant result that, assuming a false null hypothesis, may be attributed to a loss of statistical power.

**Summary**

In summary, in all three examples, the researchers applied a Bonferroni correction to adjust $α_{Constituent}$ from 0.050 to 0.025 in order to control $α_{Joint}$ at 0.050. In all three studies, the researchers found a significant result in which $p < 0.025$ and a nonsignificant result in which $0.025 < p < 0.050$. This pattern of results would allow the researchers to either (a) reject the intersection null hypothesis on the grounds that at least one test was significant using an adjusted $α_{Constituent}$ of 0.025 or (b) reject both individual null hypotheses on the grounds that both tests were significant using an unadjusted $α_{Individual}$ of 0.050. Instead, in all three cases, the researchers followed a fallacious hybrid approach in which they used an $α_{Constituent}$ of 0.025 to (a) reject one of the two individual null hypotheses and (b) fail to reject the other one. This hybrid approach is logically inconsistent with the use of a multiple testing correction. Furthermore, assuming that the null hypotheses were false, the researcher's nonsignificant results can be attributed to a loss of statistical power caused by their inconsistent corrections: If they had used an unadjusted $α_{Individual}$ of 0.050, then they would have decided that *both* of their tests yielded significant results. Their nonsignificant results also had implications for their substantive conclusions. In two of the three cases, the researchers described their results as providing only *partial* support for their hypotheses (Clemens & Grolig, 2023; Prem et al., 2021, Study 1). In fact, whichever way the results are interpreted, they provided *full* support for the hypotheses: The single significant result at the 0.025 $α_{Constituent}$ level was sufficient to reject the entire intersection null hypothesis, and the two significant results at the 0.050 $α_{Individual}$ level were sufficient to reject each of the two individual null hypotheses.

I restricted my three examples to studies published in psychology journals that used a Bonferroni correction involving two simple tests in which one test yielded a significant result at the corrected alpha level and the other yielded a nonsignificant result. Nonetheless, inconsistent corrections may also be observed among nonpsychology studies that use other family-based alpha correction approaches and larger families of tests.

## Moving Away from Statisticism

In my view, *statisticism* refers to an overgeneralization of abstract statistical principles at the expense of context-specific nuance and caveats (e.g., Boring, 1919; Brower, 1949). Statisticism may help to explain the unthinking statistical ritualism that has been noted by some commentators (Davidson, 2018; Gigerenzer, 2004, 2018; Proulx & Morey, 2021). In the area of significance testing, this ritualism may lead researchers to (a) preregister analyses and demote exploratory analyses as "tentative," even when significance tests retain their validity in non-preregistered, exploratory situations (Devezer et al., 2021; Rubin, 2017, 2020a); (b) use a conventional alpha level when an alternative unconventional alpha level is more appropriate (Lakens et al., 2018); (c) use a two-sided test when a one-sided test is more consistent with one's statistical inference (Georgiev, 2018; Rubin, 2022); (d) conduct an a priori power analysis when there is no clear basis for an effect size estimate and a sensitivity power analysis is more appropriate (Lakens, 2022; Perugini et al., 2018); and (e) follow a Neyman-Pearson interpretation when a Fisherian interpretation is more appropriate (Hurlbert & Lombardi, 2009; Rubin, 2020b).

Perhaps fuelled by concerns about statistical rigour following the replication crisis, statisticism may also help to explain a renewed promulgation of the alpha adjustment ritual. Inconsistent multiple testing corrections then follow as an overgeneralized response to a fairly limited problem.

To move away from statisticism, we need to adopt a more nuanced, context-sensitive approach that pays closer attention to the specific statistical inferences that researchers actually make. In the case of multiple testing corrections, this more nuanced approach includes the abandonment of the alpha adjustment ritual and the adoption of an inference-based perspective that advocates an alpha adjustment in the case of inferences about intersection null hypotheses but not in the case of inferences about individual null hypotheses.

Inconsistent Multiple Testing Corrections	14Frane, A. V. (2015). Are per-family type I error rates relevant in social and behavioral science? *Journal of Modern Applied Statistical Methods, 14*(1), Article 5. http://dx.doi.org/10.22237/jmasm/1430453040

Frane, A. V. (2019). Misguided opposition to multiplicity adjustment remains a problem. *Journal of Modern Applied Statistical Methods, 18*(2), eP2836. http://dx.doi.org/10.22237/jmasm/1556669400

García-Pérez, M. A. (2023). Use and misuse of corrections for multiple testing. *Methods in Psychology*, *8*, Article 100120. https://doi.org/10.1016/j.metip.2023.100120

Georgiev, G. Z. (2018, August 6). Directional claims require directional (statistical) hypotheses. *One-sided.org.* https://www.onesided.org/articles/directional-claims-require-directional-hypotheses.php

Gigerenzer, G. (2004). Mindless statistics. *The Journal of Socio-Economics, 33*(5), 587-606. https://doi.org/10.1016/j.socec.2004.09.033

Gigerenzer, G. (2018). Statistical rituals: The replication delusion and how we got there. *Advances in Methods and Practices in Psychological Science, 1*(2), 198-218. https://doi.org/10.1177/2515245918771329

Greenland, S. (2021). Analysis goals, error-cost sensitivity, and analysis hacking: Essential considerations in hypothesis testing and multiple comparisons. *Paediatric and Perinatal Epidemiology, 35,* 8-23. https://doi.org/10.1111/ppe.12711

Hewes, D. E. (2003). Methods as tools. *Human Communication Research, 29,* 448-454. https://doi.org/10.1111/j.1468-2958.2003.tb00847.x

Hitchcock, C., & Sober, E. (2004). Prediction versus accommodation and the risk of overfitting. *British Journal for the Philosophy of Science, 55*(1), 1-34. https://doi.org/10.1093/bjps/55.1.1

Hochberg, Y., & Tamrane, A. C. (1987). *Multiple comparison procedures.* Wiley. https://www.nature.com/srep/author-instructions/submission-guidelines

Hurlbert, S. H., & Lombardi, C. M. (2009). Final collapse of the Neyman-Pearson decision theoretic framework and rise of the neoFisherian. *Annales Zoologici Fennici, 46*(5), 311–349. https://doi.org/10.5735/086.046.0501

Hurlbert, S. H., & Lombardi, C. M. (2012). Lopsided reasoning on lopsided tests and multiple comparisons. *Australian & New Zealand Journal of Statistics, 54*(1), 23-42. https://doi.org/10.1111/j.1467-842X.2012.00652.x

Janssen, E. M., van Gog, T., van de Groep, L., de Lange, A. J., Knopper, R. L., Onan, E., ... & de Bruin, A. B. (2023). The role of mental effort in students' perceptions of the effectiveness of interleaved and blocked study strategies and their willingness to use them. *Educational Psychology Review, 35*(3), Article 85. https://doi.org/10.1007/s10648-023-09797-3

Kim, K., Zakharkin, S. O., Loraine, A., & Allison, D. B. (2004). Picking the most likely candidates for further development: Novel intersection-union tests for addressing multi-component hypotheses in comparative genomics. *Proceedings of the American Statistical Association, ASA Section on ENAR Spring Meeting* (pp. 1396-1402). http://www.uab.edu/cngi/pdf/2004/JSM%202004%20-IUTs%20Kim%20et%20al.pdf

Kuzon, W., Urbanchek, M., & McCabe, S. (1996). The seven deadly sins of statistical analysis. *Annals of Plastic Surgery, 37,* 265-272. http://dx.doi.org/10.1016%2FS0278-2391(97)90377-3

*Peer review*: This article been peer reviewed at *Methods in Psychology*.

*Acknowledgements*: I am grateful to Vinay Tummarakota for discussions that led to my explanation of Confusion IV.

*Funding:* I declare no funding sources.

*Conflict of interest:* I declare no conflict of interest.

*Biography*: I am a professor of psychology at Durham University, UK. For further information about my work in this area, please visit https://sites.google.com/site/markrubinsocialpsychresearch/replication-crisis

*Correspondence:* Correspondence should be addressed to Mark Rubin at the Department of Psychology, Durham University, South Road, Durham, DH1 3LE, UK. E-mail: Mark-Rubin@outlook.com